\documentclass[]{revtex4}
\usepackage{amsmath}
\usepackage{gensymb}
\usepackage{amssymb}
\usepackage{graphicx}
\usepackage{bm}

\def\cD{{\cal D}}
\def\cD{{ D}}

\def\eqnn#1{Eq.~(\ref{eq:#1})}

\def\figno#1{Fig.~\ref{fig:#1}}

\def\Im{{\rm \,Im}}

\def\vev#1{\left\langle #1\right\rangle}
\def\vev#1{\langle #1\rangle}
\def\Im{{\rm \,Im\,}}
\def\Re{{\rm \,Re\,}}
\def\kb{k_{\scriptscriptstyle\rm B}}

\def\NA{N\!A}

\begin{document}
\title{
  Thermal interface fluctuations of liquids and viscoelastic materials
}
\author{Kenichiro Aoki and Takahisa   Mitsui}
\affiliation{Research and Education Center for Natural Sciences and
  Dept. of Physics, Hiyoshi, Keio University, Yokohama 223--8521,
  Japan}
\begin{abstract}
    Spectra of thermal fluctuations of a wide range of interfaces,
     from liquid/air, viscoelastic material/air, liquid/liquid, to
    liquid/viscoelastic material interfaces, were measured over
    100\,Hz to 10\,MHz frequency range.  The obtained spectra were
    compared with the fluctuation theory of interfaces, and found to
    be mostly in quite good agreement, 
    when the
    theory was generalized to apply to thermal fluctuations of
    liquid/viscoelastic material interfaces.  The spectra were
    measured using a system that combines light reflection,
    statistical noise reduction through averaged correlations, and
    confocal microscopy. It requires only a small area of the
    interface ($\sim1\,\mu$m$^2$) , relatively short times for
    measurements ($\lesssim$few\,min), and can also be applied to
    highly viscous materials.
\end{abstract}
 \maketitle
 \section{Introduction}
 \label{sec:intro}
Interfaces between various kinds of materials, such as gas, liquid,
viscoelastic material, contain the physics of matter in the bulk, as
well as the physics of the matter specific to the interfaces, making
their study of fundamental interest\cite{intPhysics}.
The behavior of matter at interfaces is of interest, and has practical
importance, also in a broad range of areas including properties of
emulsions, lubrication, oil recovery and decontamination, as well as
biological applications\cite{intGeneral}.  One effective approach in
investigating the dynamics of interfaces is to study their thermal
fluctuations non-invasively.  While large thermal fluctuations in the critical
regime have been seen\cite{Lauter,Giglio}, generically the
motions are at atomic scales, and measuring their properties with
precision remains to be a challenging subject.  Both for fundamental
physics, and for practical applications, we believe it is important to
investigate spontaneous fluctuations at interfaces in detail, and to
understand the validity of fundamental theories that have been
proposed to describe their behavior.

In this work, we directly measure the spectra of thermal fluctuations
of various types of interfaces at room temperature, including
liquid/liquid, and liquid/viscoelastic matter interfaces.  Previously,
thermal fluctuations of liquid/air and liquid/liquid interfaces, often
called ``ripplons''\cite{ripplon}, have been observed in light
scattering experiments, using propagating waves on the interfaces
effectively as traveling
gratings\cite{ripplonExp,Domingo,Sauer,Sawada94}.  Thermal
fluctuations of liquid/viscoelastic material interfaces seem not to
have been observed previously.  Using light scattering to observe
thermal fluctuations is difficult for interfaces involving liquids
with large viscosity, since the waves dissipate\cite{SakaiVisc}.  The
theory of thermal fluctuations of interfaces was worked out from first
principles some time ago\cite{Levich,Bouchiat,HM,Meunier,HPP}, yet it
had not been possible to experimentally examine their behavior in
detail.  In this work, the fluctuations are observed using light
reflection, with the interface effectively acting as an optical
lever\cite{lever,MitsuiOnion,Tay,MA1}, so that the method does not
rely on the existence of waves that act as gratings.  This allows us
to measure properties of thermal fluctuations of previously
inaccessible interfaces with precision, over a wide frequency range,
including those involving highly viscous materials. The obtained
fluctuation spectra were analyzed in view of the theory, which was
generalized to apply also to liquid/viscoelastic material interfaces.
The results mostly show good agreement with the theory, but aspects
that need to be further investigated remain.

\section{Experimental concept and scheme}
\label{sec:exp}
\begin{figure}[htbp]\centering
     \includegraphics[width=80mm,clip=true]{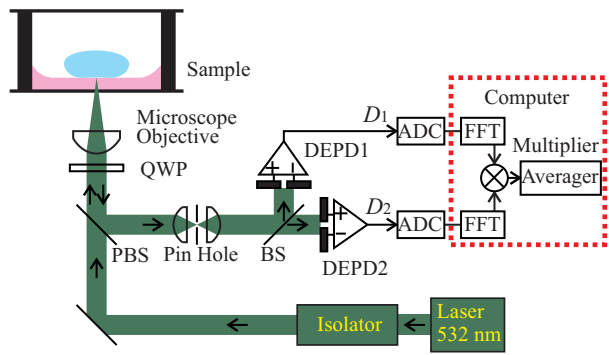}
     \caption{The setup of the measurement system: Light from the
       laser source (wavelength 532\,nm) is stabilized using an
       isolator, and is focused on the sample interface between two
       media (blue, pink). Using a quarter-wave plate (QWP), the light
       from the sample is reflected at the polarized beam splitter
       (PBS). A pin hole is used for confocal microscopy.  The
       reflected light is partitioned in two by a beam splitter (BS)
       and the light beams are detected by dual-element photodiodes
       (DEPD1,2).  The measured photocurrents are digitized by
       analog-to-digital converters (ADC), and processed on a
       computer.  }
\label{fig:setup}
\end{figure}
To investigate the behavior of interfaces precisely, the basic
experimental concept we adopt is to measure the fluctuations in the
average inclination of the interface in the light beam, which acts as a
partial mirror.  This basic principle is realized in the experimental
setup shown in \figno{setup}: A light beam is focused on the interface
between two media, and is partially reflected. At any instant, the
average direction of the reflected light deviates slightly from the
incoming beam direction, due to interface fluctuations. These
inclination fluctuations are measured through the photocurrent
differential in a dual-element photodiode (DEPD), and their power
spectrum is computed.  Confocal microscopy is used to select {\it
  only} the fluctuations at the interface under investigation.
The measurement method employed can be described as dynamic light
reflection, and compared to dynamic light scattering, the signal is
stronger for a given incoming light, so that a far weaker light source can
be used. 

For precision measurements, it is imperative to reduce the shot-noise,
which appears as $2eI$ in the photocurrent ($e$ is the electron
charge, and $I$ is the photocurrent). The shot-noise originates in the photon
nature of light, and can potentially overwhelm the signal.  One
option is to increase the light power, which will, in general, affect
the sample significantly, and in the problem at hand, it is not
possible to increase the light powers to levels where the spectra can
be recovered to the desired accuracy without doing so. In this work,
we reduced the shot-noise statistically, as follows: The light
reflected from the interface is split into two beams, whose
(time-dependent) powers are measured independently by photodetectors
as $\cD_{1,2}$.  These measurements, $\cD_j=S+N_j$, contain both the
signal, $S$, and the noise, $N_j$, such as shot-noise, that occurs
independently in the photocurrents. By {\it repetitively} taking the
correlation of the Fourier transform of these measurements, and
averaging over them, the desired spectrum is obtained in the limit of
infinite number of averagings, $ \vev{\overline{\tilde \cD_1}\tilde
  \cD_2} \longrightarrow\vev{|\tilde S|^2}$.  Here, $\vev{\cdots}$
denotes ensemble averaged results, and tilde denotes the Fourier
transform. Any noise that is decorrelated in the two measurements,
including the shot-noise and amplifier noise, is reduced statistically
in the correlation $\vev{\overline{\tilde N_1}\tilde N_2}$, by a
factor $1/\sqrt{{\cal T} \Delta f}$, $\Delta f$ being the frequency
resolution, ${\cal T}$ the total measurement time. This reduction is
statistical, and is consistent with the statistical fluctuations of
the photons, since the total number of observed photons has increased
by a factor of ${{\cal T} \Delta f}$, the number of averagings. Taking
the correlation of independent measurements here is crucial, since
without it, the shot-noise is not reduced.  This principle for
reducing the shot-noise has been used to achieve factors of $10^{-3}$
to $10^{-5}$ reduction, in the measurements of surface thermal
fluctuations of fluids\cite{MA1,MA2}, spontaneous noise in atomic
vapor\cite{MA3,MA4}, and reflectance fluctuations\cite{MA5}.

Technical details of the setup (\figno{setup}) are as follows: Light
from a laser source (wavelength 532\,nm), stabilized by an isolator,
is focused on the interface between two materials close to the
diffraction limit, using a microscope objective
with a correction mechanism for the aberration caused
by the material below the interface, and a numerical aperture ($\NA$)
of 0.6.
The materials were contained in a cylinder with a inner
diameter of 12.2\,mm. The total light powers applied on the interfaces
varied from $0.4$ to $1.0$\,mW. A quarter-wave plate is included in the
light path to reflect the light coming from the sample interface at
the polarized beam splitter. Using confocal microscopy, a pin
hole is placed in the reflected light beam path to select out the
light reflected from the interface.  The light reflected at the sample
interface is then split in two using a beam splitter, and the light
beam powers are measured using dual-element photodiodes (DEPD's).
The two reflected light beams are so directed that the light powers in
the two photodiodes in each of DEPD1,2 are identical, apart from
noise. Fluctuations cause the interface to act as an optical
lever\cite{lever}, causing differences in the photocurrents from DEPD.
These differential photocurrents are fed through an analog-to-digital
converter into a computer,
which computes their Fourier transforms (FFT), correlations, and 
averages. Measurement times for the spectra obtained below varied from
15 seconds to 7 minutes, depending on the reflectance of the
interface, the magnitude of the fluctuations, and the desired precision.
The average inclination, $\theta$, of the surface within the beam spot
is related to the photocurrents in the two elements in DEPD's as
$(I_{j1}-I_{j2})/(I_{j1}+I_{j2})=2\theta/\NA$, where $I_{j1},I_{j2}$
are the photocurrents from the two elements in DEPD$j\ (j=1,2)$.
Since the light beam diameter is slightly smaller than that of the microscope objective,
$\NA$ can effectively be smaller than its catalog value.
 The
power spectrum is computed in the standard fashion\cite{SpectrumRef},
as $S(\omega)=\vev{|\tilde \theta|^2}/{\cal T}={\rm const.}\times
\vev{\overline{\tilde D_1}\tilde D_2}/{\cal T}$, where
$\omega$ is the angular frequency.

\section{Interfaces of liquids}
\label{sec:liquid}
Theoretically, the fluctuation spectrum of the inclinations of the
surface is
\begin{equation}
    \label{eq:specInc}
    S(f)=\int_0^\infty dq\,q^3e^{-w^{2}q^2/4}P(q,\omega)\quad,    
\end{equation}
where $P(q,\omega)$ is the spectral function for the interface
fluctuations, and $w$ is the beam spot radius, and $q$ denotes the
wave number \cite{MA1,AM1}. Since the observed inclination of the
surface is averaged within the beam spot, fluctuations with shorter
wavelengths are effectively cutoff by a Gaussian factor.  The spectral
function for the thermal interface fluctuations between two fluids,
described by densities $\rho_{1,2}$, shear viscosities $\eta_{1,2}$,
and interface tension, $\sigma$ has been derived in \cite{HM,Meunier};
\begin{equation}
    \label{eq:Pqw}
    P(q,\omega)=-{\kb T\over \pi\omega} {q\over \rho_1+\rho_2}\Im F^{-1},\quad
    F\equiv {1\over\tau_0^{2}}\left[y+s^2+2s{(\eta_1+\eta_2m_2)(\eta_2+\eta_1m_1)
      \over (\eta_1+\eta_2)\left[\eta_1(m_1+1)+\eta_2(m_2+1)\right]}\right]\quad,
\end{equation}
\begin{equation}
    \label{eq:PqwDef}
 \tau_0={\rho_1+\rho_2\over 2(\eta_1+\eta_2)q^2},\quad 
 y={\sigma(\rho_1+\rho_2)\over 4(\eta_1+\eta_2)^2q},\quad
 s= i\omega\tau_0,\quad
m_j= \sqrt{1+i{\rho_j\omega\over\eta_j q^2}},\quad (j=1,2)
    \quad,
\end{equation}
where  we have
also made use of the expressions in
\cite{Loudon,Loudon2}.
Gravity affects interface fluctuations in the region where the
wavelength is larger than $2\pi\sqrt{\sigma/(|\rho_1-\rho_2| g)}$,
where $g$ is the gravitational acceleration. These fluctuations have
wavelengths of few centimeters or larger, which are suppressed by the
experimental geometry, so will not be considered below.
The spectral function generalizes that of the surface (liquid/air
interface)\cite{Bouchiat}, to which it reduces to in the limit
$\rho_1,\eta_1\rightarrow0$, assuming medium 1 is air. The spectral
function for the liquid surface (liquid/air interface) was generalized
to surface fluctuations of polymer solutions, by incorporating the
contributions from the polymer network effectively into the fluid
viscosity as\cite{HPP}
\begin{equation}
    \label{eq:visc}
    \eta=\eta_0+{G\tau\over(1+ i\omega\tau)}     
\end{equation}
Here, $\eta_0$ is the viscosity of the solvent, $G$ is the shear
modulus, and $\tau$ is the stress relaxation time. Polymer gel surface
fluctuation spectral function is obtained in the limit,
$\tau\rightarrow\infty$.

\begin{figure}[htbp]\centering
    \includegraphics[width=80mm,clip=true]{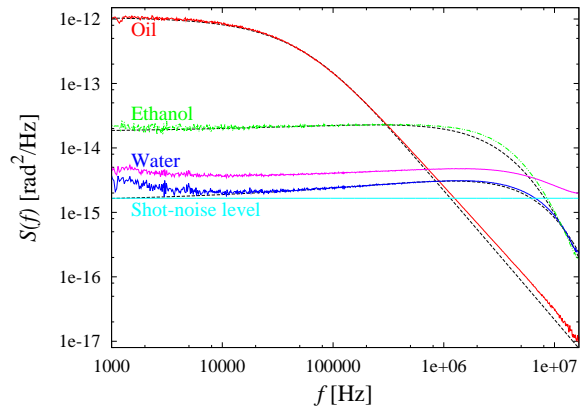}
    \caption{Measured surface fluctuation spectra for
      water (blue), ethanol (green), and oil (red).  The corresponding
      spectra derived from theory are also shown (black dashes), which
      agree with the results well. For
      the water surface, the spectrum measured without using averaged 
      correlations is also shown (magenta), along with the
      corresponding shot-noise level (cyan).  Without using correlations, the 
    spectrum contains the shot-noise, and differs visibly from the
    spectrum without the shot-noise.}
\label{fig:surf}
\end{figure}
In \figno{surf}, the measured surface inclination fluctuations for
water, ethanol and oil (Olympus immersion oil AX9602, here and below)
are shown, along with their theoretical spectra, computed from
\eqnn{specInc},(\ref{eq:Pqw}). Density, viscosity, surface tension, 
temperature of the fluids were $(\rho_2\, {\rm[kg/m^3]},\eta_2\,{\rm
  [Pa\cdot s]},\sigma\,{\rm[N/m]}, t\,[\rm \degree C])
=(9.98\times10^2,7.26\times10^{-2},9.58\times10^{-2},22),
(7.89\times10^2,2.25\times10^{-2},1.15\times10^{-3},22),
(9.20\times10^2,3.00\times10^{-2},0.124,24)$, respectively\cite{Prop},
and
$w=0.6\,\mu$m.  The theory and the experiment are seen to agree
well. These spectra are essentially the same as those measured in
\cite{MA1}, except that the light was shone from inside the liquid, in
contrast to the previous work, in which the light was shone from
outside the liquid. 
 For the water surface, the
fluctuation spectrum
obtained by averaging $\vev{|\tilde D_j|^2} \ (j=1,2)$, without using
correlations, is also shown, along with the corresponding shot-noise level.
Clearly, the spectrum differs significantly from that with the
shot-noise removed. The noise level in an interface inclination
fluctuation spectrum originating in the shot-noise,  which we
call the ``shot-noise level'' for brevity, is $\NA^2 e/(4I n^2)$,
where $n$ is the index of refraction of the lower medium in
\figno{setup}, and $I$ is the photocurrent per photodiode.  
The overall magnitudes of the spectra were normalized by their
corresponding theoretical spectra.  
For all the spectra shown in
\figno{surf} and below, this shot-noise level computed using the
photocurrent agreed with the observed shot-noise levels within
experimental uncertainties.

Thermal surface fluctuations of liquids, often called ``ripplons''\cite{ripplon},
have previously been observed using surface light scattering, using
the surface waves as gratings\cite{ripplonExp}. In these measurements, the
spectral function for particular wave numbers are obtained. In
contrast, in our measurements, the spectrum integrated over the wave
number up to the scale of inverse of the beam spot radius is
obtained. With previous surface scattering methods, surface
fluctuations of highly viscous liquids were difficult to
measure\cite{SakaiVisc}, whereas there is no additional effort
involved for such liquids with our method.

\begin{figure}[htbp]\centering
    \includegraphics[width=80mm,clip=true]{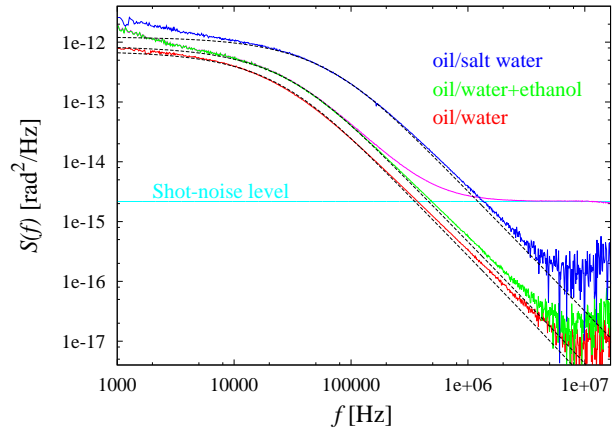}
    \caption{ Measured inclination fluctuation spectra for interfaces
      of oil with water (red), salt water (concentration 16\,\%,
      blue), and water-ethanol mixture (20\,\% ethanol, green).  The
      corresponding spectra derived from theory are also shown (black
      dashes), which are quite consistent with the experimental
      results. The spectrum measured without using averaged
      correlations is shown for the oil-water interface (magenta), with
      the corresponding shot-noise level (cyan).  Higher frequency
      fluctuations are dominated by the shot-noise, unless it is
      removed. 
}
\label{fig:oilWater}
\end{figure}
In \figno{oilWater}, the thermal inclination fluctuation spectra for
interfaces of oil with
water, salt water (concentration 16\,\%), and water-ethanol mixture
(20\,\% ethanol), are shown.
Corresponding theoretical spectra obtained from \eqnn{Pqw} are also
shown for each spectrum, which agree with the experimental results
well.
The density and viscosity of oil are common to these measurements,
$(\rho_1\, \rm[kg/m^3],\eta_1\,[Pa\cdot s])=(920,0.124)$, and for the
water solutions, interface tension, beam radius, and the temperature
were $(\rho_2\, {\rm[kg/m^3]},\eta_2\,{\rm [Pa\cdot
  s]},\sigma\,{\rm[N/m]},w\,{\rm [\mu m]},t\,[\rm \degree
C])=(998,9.58\times10^{-4},0.026,1.2,22),
(1121,1.37\times10^{-3},0.030,0.525,24),(973,1.72\times10^{-3},0.026,1.0,22)
$, respectively. 
Here, we adopted the known properties for the bulk
properties of water solutions\cite{Prop,SaltWater,Ethanol}.
The interface tension values obtained here seem consistent with the
theory, and previous measurements\cite{Fowkes,FisherFood,KimBurgess}.
%
The shape of the fluctuation spectra for these interfaces are similar
to each other,  and also to that of the oil/air interface in \figno{surf}.
Theoretically, the strong viscosity of oil dominates
the spectra, and they should not have visible dependence on the
concentration, unless the interface tension is changed, which is
consistent with our measurements.  The concentrations of salt and
ethanol were systematically varied up to the levels seen in
\figno{oilWater}, but differences in interface fluctuation spectra due
to the concentration were not observed, which is consistent with
previous interface tension measurements\cite{Gaonkar}.  The
differences observed in \figno{oilWater} can be attributed to the
differences in the values of $w$.  Theoretically, from the spectral
function \eqnn{Pqw}, the properties of the spectra are governed in
large part by the most viscous fluid, when the viscosities of the
fluids are widely different, as is the case here.
The shot-noise level for the oil/water-ethanol mixture interface is
additionally indicated in \figno{oilWater}, and it can be seen that
fluctuations at higher frequencies are buried below the shot-noise, if
this is not reduced.
Liquid/liquid interface fluctuations involving such highly viscous
fluids have not been previously measured, to our knowledge, due to the
absence of persistent waves, and the small reflectance of the
interface.  The frequency regions around the peak in the spectral
functions have been observed previously for less viscous liquids
\cite{Domingo,Sauer,Sawada94}, but fluctuations over a wide frequency
range, especially including the high frequencies have not been
previously observed.

\section{Interfaces of liquids and viscoelastic materials}
\label{sec:viscoelastic}
\begin{figure}[htbp]\centering
    \includegraphics[width=80mm,clip=true]{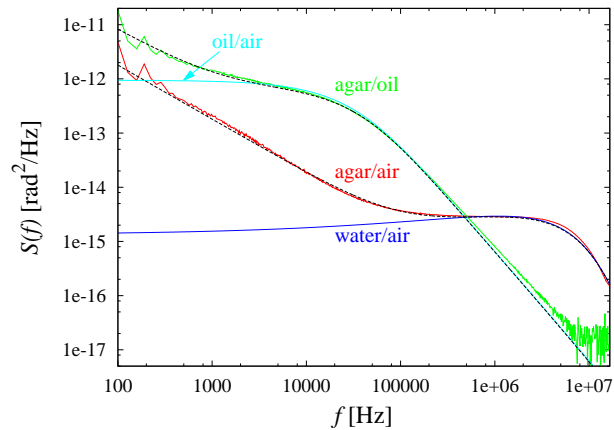}
    \caption{Observed inclination fluctuation spectra of agar/air
      (red) and agar/oil (green) interfaces. The corresponding
      theoretical spectra (black dashes) agree quite well with the
      observed spectra.  Theoretical spectrum for water/air interface
      fluctuations (blue) agree well with the agar/air interface
      fluctuations at higher frequencies ($f\gtrsim10^6$\,Hz).  The
      theoretical oil/air spectrum (cyan) matches the observed agar/oil
      spectrum well for higher frequencies, but not as well at lower
      frequencies ($f\lesssim10^4$\,Hz). }
\label{fig:agar}
\end{figure}
\label{sec:viscoelastic}
\begin{figure}[htbp]\centering
    \includegraphics[width=80mm,clip=true]{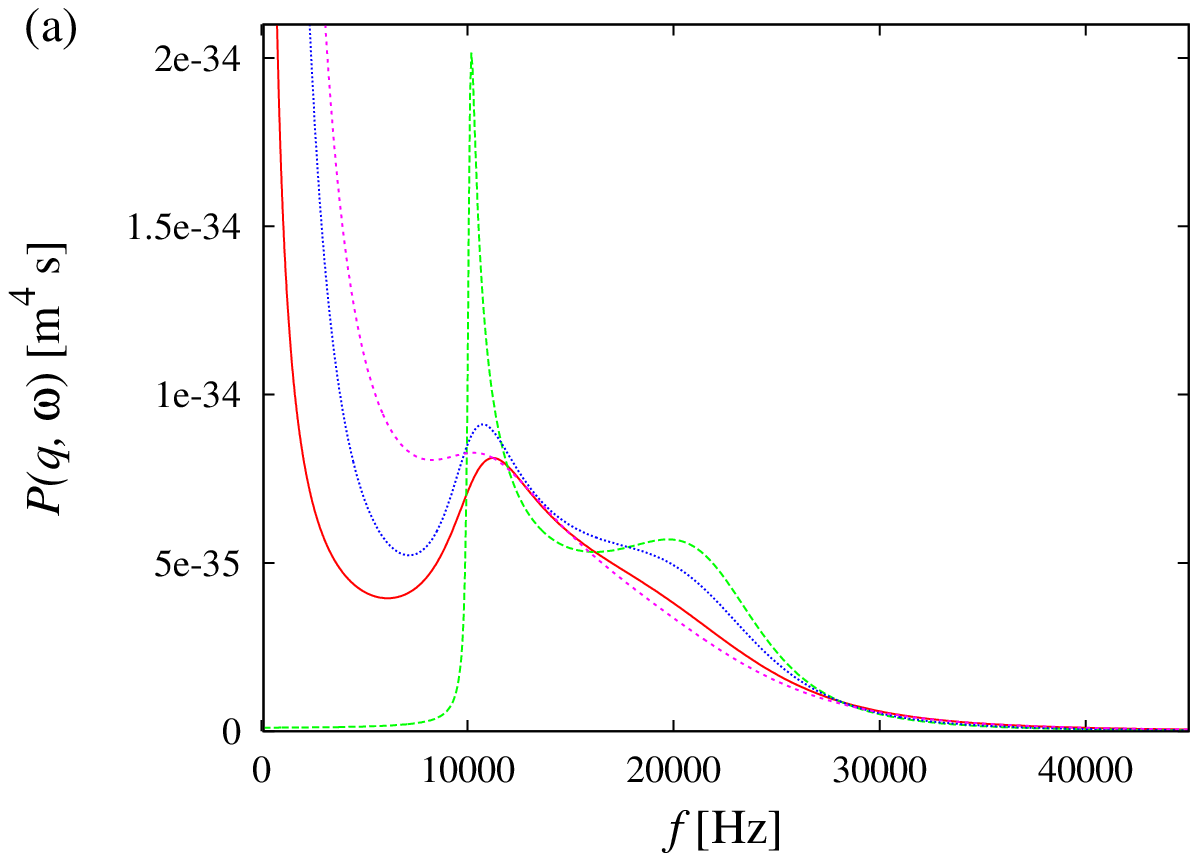}
    \includegraphics[width=80mm,clip=true]{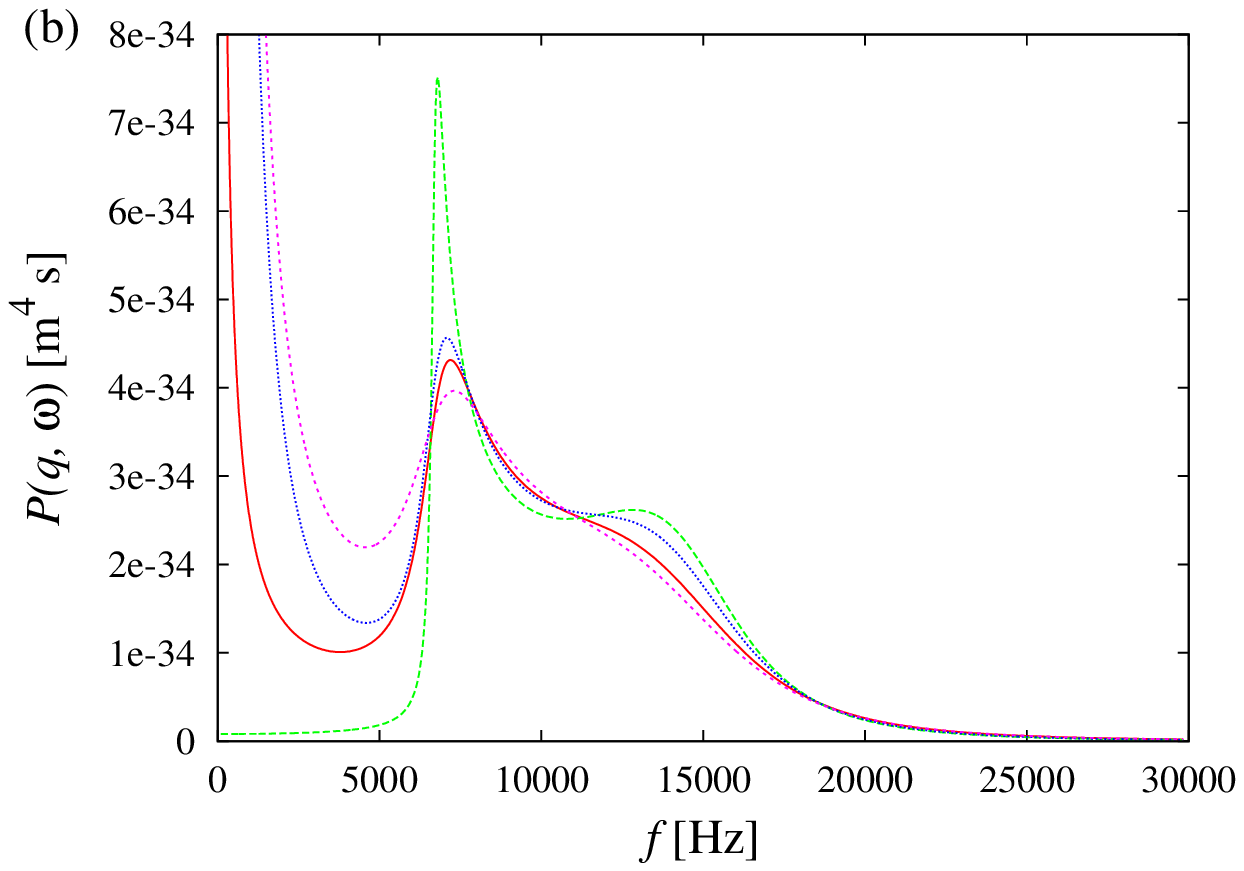}
    \caption{Comparison of the behaviors of the surface thermal
      fluctuation spectral function, $P(q,\omega)$, for polymer gel
      with loss modulus (red, solid), without (green, long dashes),
      polymer sol with loss modulus (blue, dots), and without
      (magenta, short dashes). The parameters used were, for (a),
      $\rho=10^3\,\rm kg/m^3,\ \sigma= 0.073\, N/m,\
      \eta=10^{-3}\,Pa\cdot s,$, $
      G=(2.5\times10^3+i1.0\times10^3)\rm \,Pa,\ \tau=5\times10^{-5}\,s$, and for (b), 
      $\rho=7.3\times10^2\,\rm kg/m^3,\ \sigma= 0.024\, N/m,\
      \eta=8\times10^{-4}\,Pa\cdot s$, $
      G=(8.0\times10^2+i1.5\times10^2)\,\rm Pa,\ \tau=1.5\times10^{-4}\,s$.
      $\Im G=0$ when loss modulus is not considered, and $\tau$ is
      infinite for gels.  $q=4\times10^4{\rm \,m^{-1}},$ and
      $t=27\,$\degree C, both for (a), and (b).  }
\label{fig:solGel}
\end{figure}
In \figno{agar}, the inclination fluctuation spectra for agar/air, and
agar/oil interfaces are shown, along with the corresponding
theoretical spectra (agarose gel is Agar, Difco 214220 2.6\,wt\,\% in
water).  The theoretical spectrum for the agar/air interface was
obtained using the formalism of \cite{HPP} in \eqnn{Pqw}, with
$G=(2.5\times10^3+i1.0\times10^3)$\,Pa, $w=0.65\,\mu$m,
$t=24\,\degree$C, the properties of water as the solvent, which
reproduces the observed spectra quite well. $G$, which is not real,
contains both the shear modulus, and the loss modulus.
At low frequencies, the leading behavior for the spectrum is
$\Re\eta/(\sigma^2w)$\cite{AM1}, so that it falls off as $f^{-1}$ when
$\Im G\not=0$, and behaves as a constant when $\Im G=0$ (or as
$f^{-2}\tau^{-1}$ for sols). Therefore, the effect of the non-zero
loss modulus is clearly visible in the spectrum above, for
$f\lesssim200\,$kHz.  Viscoelastic material/air interface fluctuations
have been observed previously with light scattering
techniques\cite{Cao91,Kikuchi91,Dorshow93,Cao95,Huang96,Monroy} ,
where a non-zero loss modulus was not considered, though viscoelastic
materials have non-zero loss moduli, in general\cite{Ferry}.  The
effect of the loss modulus is significant in the region, $\Im
G/(\eta_0\omega)\gtrsim1$. On the other hand, at relatively high
frequencies, and relatively high wave numbers associated to it through
the dispersion relation \eqnn{Pqw}, the effect is small, which can
explain why it had not been noticed previously. 
This is illustrated in \figno{solGel}, where the theoretical spectral
behaviors of polymer gels and sols, with or without the loss modulus
are compared, for parameters pertinent to our experiment in (a), and
for parameters in the typical range for past measurements, such as
those of polyisobutylene decane
solutions\cite{Cao91,Dorshow93,Huang96}, in (b). It can be seen that
polymer gel with the loss modulus, and polymer sol, both introduce
dissipation to the dispersion relation peak in the gel spectral
function, leading to similar behaviors around it. However, the
behaviors at lower frequencies are qualitatively different, constant
for gel without the loss modulus, $f^{-1}$ with it, and $f^{-2}$ for
sols.
In \figno{agar}, it can also be seen that the high frequency
fluctuations ($f\gtrsim3\times10^5\,$Hz) are well described by
properties of the solvent, water.
The values of shear and loss modulus seem to be consistent with those
measured before\cite{Monroy,Kikuchi94}, and also with those
measured by other methods\cite{Tokita,Fujii}.

For the theory of thermal fluctuations of the agar/oil interface, we
generalize the interface fluctuation formula \eqnn{Pqw}
\cite{HM,Meunier}, by incorporating the polymer network contributions,
as was done for the surface\cite{HPP}. While this generalization has
not been presented before, it is a natural one that can be derived
much in the same manner\cite{Meunier,HM,HPP}. Applying this theory,
with the value of $G$ used for the agar/air interface, $w=0.9\,\mu \rm
m$, and $t=24\degree C$, the computed spectrum matches excellently
with the observed spectrum (\figno{agar}). For comparison, the
theoretical surface fluctuation spectrum of oil with the same surface
tension is shown, and it can be seen that a qualitative difference
exists for low frequencies ($f\lesssim10^4\,$Hz), while the higher
frequency fluctuations are dominated by the properties of
oil. Interfaces fluctuations between liquids and viscoelastic
materials have not been observed previously.

\begin{figure}[htbp]\centering
    \includegraphics[width=80mm,clip=true]{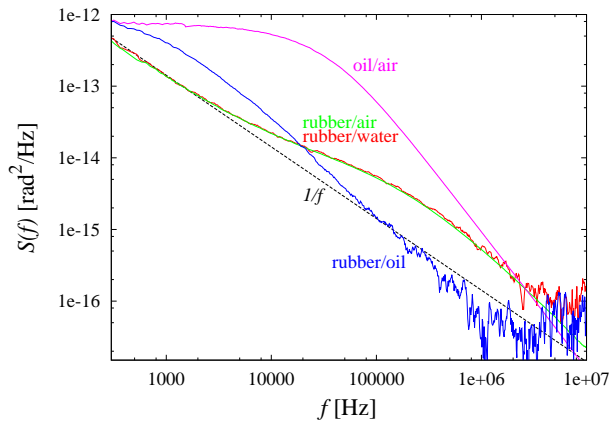}
    \caption{Observed inclination fluctuation spectra for the
      interfaces of silicone rubber with
      air (green), water (red), and oil (blue). For reference, the
      spectrum for the oil/air interface is also shown (magenta),
      along with its corresponding $1/f$ spectrum (black,
      dashes). 
    }
\label{fig:si}
\end{figure}
In \figno{si}, thermal inclination fluctuation spectra of silicone
rubber (Tigers Polymer, SR 0.2\,t, Japan) interface with air, water,
and oil are shown ($w=0.6,\mu$m, $t=24\,$\degree C), along with that
of the oil/air interface. Thermal fluctuations at interfaces of
elastic materials and liquids seem not to have been observed
previously.
Since rubber is an elastic material with no solvent, its
interface fluctuations can be obtained by taking the limit of
viscosity going to zero in \eqnn{Pqw}, which we shall use here. The
spectral function computed this way differs from that obtained
by other methods\cite{Saulson90,Levin,Braginsky99} by a dimensionless
prefactor of order 1. 
Interestingly, rubber/air and rubber/oil interface spectra differ
significantly, and unlike the agar/oil interface spectrum in
\figno{agar}, rubber/oil spectrum seems unrelated to the oil surface
fluctuation spectrum, which is also shown for contrast.
For rubber, if we assume that $G$ is frequency independent,
the integrated spectrum behaves as $1/f $, in this frequency range. In
\figno{si}, we compared the measured thermal fluctuation spectrum
to the theoretical one computed with
$G=(3.5\times10^6+i3.5\times10^5)\,$Pa, which captures the
overall spectrum but not the deviations from the $1/f$ behavior.  This
is not surprising since the loss modulus of rubber is frequency
dependent, and the value of $G$ seems consistent with the known properties
\cite{Ferry,Sperling}.  The thermal fluctuation spectrum
of rubber/water interface is essentially indistinguishable
from that of rubber/air interface in \figno{si}, and this is
quite consistent with the spectrum formula applied to these
interfaces. It is seen in \figno{si} that the thermal fluctuation
spectrum for the rubber/oil interface differs qualitatively
from that of rubber/air, or water interfaces.
To understand this, the frequency dependence of the shear
and loss moduli needs to be considered.

In this work, we measured the thermal fluctuation spectra of various
types of interfaces; liquid/air, viscoelastic material/air,
liquid/liquid, and liquid/viscoelastic material.  The corresponding
inclination fluctuation spectra were computed by combining the
fluctuation theory of interfaces\cite{Meunier,HM} with that of
surfaces of viscoelastic materials\cite{HPP}, and generalizing
them. The theoretically computed spectra were, for the most part,
found to be in excellent agreement with the experimentally observed
spectra.  We find it intriguing that the elegant simple formalism of
\cite{Levich,Bouchiat,HM,Meunier,HPP} can describe the whole spectrum
of a wide range of interface fluctuations quite well, when the loss
modulus is also considered.  For the interface fluctuation spectra
involving rubber, it was found that the frequency dependence of the
elastic constants of the material was perhaps necessary to explain the
spectra fully. We believe that the measurement system enables direct
measurements of interface phenomena that were previously inaccessible,
leading to a better understanding of the underlying fundamental
theories.  This measurement system we developed for the thermal
interface fluctuations combines light reflection measurement, noise
reduction through averaged correlations, and confocal microscopy.  It
is applicable to a wide variety of interfaces, as seen in this work.
Some features of this system is that it can be applied to obtain
fluctuation spectra of interfaces involving highly dissipative
materials, requires only a small area (beam spot radius
$\lesssim1\,\mu$m) , and relatively low light powers ($\simeq1\,$mW).
It takes a short time to take a measurement (typically under a few minutes),
which is determined by the desired precision.
The  system might also be suited for non-invasive  measurements
in biological and medical applications.
Light reflection measurements of fluctuations, while seemingly similar
to the more standard light scattering measurements, can  also yield
information regarding small wave number and low frequency behavior,
complementing them.

\section*{Acknowledgments}
\label{sec:ack}
K.A.  was supported in part by the Grant--in--Aid for Scientific
Research (Grant No.~15K05217) from the Japan Society for the Promotion
of Science, and a grant from Keio University.

\end{document}